\documentclass[a4paper,11pt]{article}
\usepackage{pos}
\usepackage{booktabs}
\usepackage{cancel}
\usepackage{tikz}
\usetikzlibrary{calc}

\def\kt{k_{\rm T}}

\def\ga{g_{\rm A}}

\def\lv{l_{\rm V}}
\def\lt{l_{\rm T}}
\def\ltt{l_{\rm \widetilde{T}}}

\def\zf{z_{\rm f}}

\begin{document}
\begin{flushright}
CERN-TH-2021-204
\end{flushright}
\title{Renormalization $\&$ improvement of the tensor operator for $N_f=3$ QCD in a $\chi$SF setup}
\ShortTitle{$\chi$SF renormalization  of  tensor operator}
\author[a]{{\includegraphics[height=2.0\baselineskip]{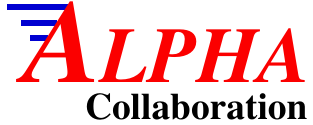}\\} Isabel~Campos~Plasencia}
\author[b]{Mattia~Dalla~Brida}
\author*[c,d]{Giulia~Maria~de~Divitiis}
\author[e]{Andrew~Lytle}
\author[f]{Mauro~Papinutto}
\author[c,d]{Ludovica~Pirelli}
\author[d]{Anastassios~Vladikas}

\affiliation[a]{Instituto de F\'{\i}sica de Cantabria IFCA-CSIC,\\
Avda. de los Castros s/n, E-39005 Santander, Spain}

\affiliation[b]{Theoretical Physics Department, CERN,\\
 CH-1211 Geneva 23, Switzerland}

\affiliation[c]{Dipartimento di Fisica, Universit\`a di Roma ``Tor Vergata'',\\
 Via della Ricerca Scientifica 1, I-00133 Rome, Italy}
  \affiliation[d]{INFN, Sezione di Tor Vergata, c/o Dipartimento di Fisica, Universit\`a di Roma ``Tor Vergata'',\\
 Via della Ricerca Scientifica 1, I-00133 Rome, Italy}

 \affiliation[e]{Department of Physics, University of Illinois at Urbana-Champaign,\\
Urbana, IL 61801, USA}

 \affiliation[f]{Dipartimento di Fisica, Universit\`a di Roma La Sapienza and INFN, Sezione di Roma, \\
Piazzale A. Moro 2, Roma, I-00185, Italy}

\emailAdd{isabel.campos@csic.es}
\emailAdd{mattia.dalla.brida@cern.ch}
\emailAdd{giulia.dedivitiis@roma2.infn.it}
\emailAdd{atlytle@illinois.edu}
\emailAdd{mauro.papinutto@roma1.infn.it}
\emailAdd{ludovica.pirelli@students.uniroma2.eu}
\emailAdd{tassos.vladikas@roma2.infn.it}

\abstract{
We present preliminary results of the non-perturbative renormalization group (RG) running  of the flavor non-singlet tensor operator.
We employ the $\chi$SF scheme for $N_f=3$ QCD using ensembles generated by the ALPHA collaboration for the computation of the quark mass running. The  $\chi$SF property of  automatic $O(a)$ improvement  prevents the $O(a)$ mixing of the correlation functions.
}
\FullConference{%
 The 38th International Symposium on Lattice Field Theory, LATTICE2021
  26th-30th July, 2021
  Zoom/Gather@Massachusetts Institute of Technology
}

%% \tableofcontents

\maketitle

%%%
\section{Flavor non-singlet tensor operator}
A non-perturbative determination of renormalisation group running between hadronic and electroweak scales  for the flavor non-singlet tensor operator
\begin{align}
T_{\mu \nu} ^a(x)=i\bar \psi (x) \; &\sigma_{\mu \nu} \; \tfrac{1}{2}\tau^a \; \psi(x)\\
&\sigma_{\mu \nu} \equiv \tfrac{i}{2} [\gamma_\mu,\gamma_\nu]\,
\end{align} 
is very interesting from both phenomenological and theoretical points of view.
The tensor enters the amplitudes of effective Hamiltonians, which describe, for example, rare heavy meson decays, neutron beta decays and possible Beyond Standard Model effects:
\begin{align}
{\cal A} = \langle f | {\cal H}_{eff} | i \rangle  = C_W(\mu)\langle f | {\cal O}(\mu) | i \rangle\\
 {\cal O}\sim (\bar l \sigma_{\mu \nu}  e)(\bar q_i \sigma_{\mu \nu}  q_j) ,\;  G_{\mu \nu} (\bar q_i \sigma_{\mu \nu}  q_j) \dots
\end{align}
Moreover, the computation of the scale dependence of the renormalization factor completes the ALPHA renormalization and improvement programme of the bilinear operators.
For $N_f=0,2$ such a study has appeared in ref.~\cite{Pena:2017hct}. For $N_f=3$, preliminary results of the RG-running in the relatively high 
energy range $2~\rm{GeV}\lesssim \mu\lesssim 128~\rm{GeV}$ have been reported in ref.~\cite{Chimirri:2019xsv}. 
$N_f=3$ renormalisation factors at different scales are also presented in ref.~\cite{Bali:2020lwx}.
%%%
\section{RG flow}
We employ a $\chi$SF setup (see \cite{Sint:2010eh,Sint:2010xy,DallaBrida:2016smt,Mainar:2016uwb,DallaBrida:2018tpn}), which is
a  mass-independent renormalization scheme. Such schemes are characterized by RG equations of the following form:
\begin{align}
  \mu \frac{\partial}{\partial \mu} T_\mathrm{R}(\mu) =\gamma  (g_\mathrm{R}(\mu)) \, &T_\mathrm{R}(\mu)\,,\qquad\qquad T_\mathrm{R}(\mu)     = Z_\mathrm{T}(\mu) T, 
\end{align}
where $g_\mathrm{R}$ is the running coupling.
The anomalous dimension $\gamma$ has the perturbative expansion
\begin{align}
  \gamma(g_\mathrm{R})  &\stackrel{g_\mathrm{R}\to0}{\sim} -{g_\mathrm{R}}^2( \gamma_0 +\gamma_1 {g_\mathrm{R}}^2 +\gamma_2 {g_\mathrm{R}}^4 + \mathcal{O}({g_\mathrm{R}}^6)) \,,
\end{align}
with a universal coefficient $\gamma_0$.
The solution $ {T} _\mathrm{R}(\mu)$ is expressed in terms of an integration constant ${T_{\scriptscriptstyle \rm RGI}}$, which is  renormalization group invariant (RGI):
\begin{align}
       {T}_{\scriptscriptstyle \rm  RGI} & =  {{T} _\mathrm{R}(\mu)} \left [ \frac{ {g_\mathrm{R}}^2(\mu) }{4 \pi} \right ]^{-\frac{\gamma_0}{2b_0}}\exp  \left \{
            -\int\limits_0^{g_\mathrm{R}(\mu)} \, { d\mathnormal{g}} \,{\left [
            \frac{\gamma(\mathnormal{g})}{\beta(\mathnormal{g})}-\frac{\gamma_0}{b_0 \mathnormal{g}} \right ]} \right \} \,.
\end{align}
It is possible to factorize the running in many evolutions between two scales:
\begin{align}{
{T_\mathrm{R}(\mu)} = 
{\frac{T_\mathrm{R}(\mu)}{T_\mathrm{R}(\mu_n)}}\;
\cdots
{\frac{T_\mathrm{R}(\mu_{2})}{T_\mathrm{R}(\mu_1)}}\;
{\frac{T_\mathrm{R}(\mu_1)}{T_{\scriptscriptstyle \rm  RGI}}}\;
{T_{\scriptscriptstyle \rm  RGI}}
\,,
}\end{align}
leading naturally to the definition of the step scaling function: 
\begin{align}
\sigma_{T}(s,u)=
\frac{T_\mathrm{R}(\mu_2)}{T_\mathrm{R}(\mu_1)}=
\frac{Z_\mathrm{T}(\mu_2)}{Z_\mathrm{T}(\mu_1)}\,,
\end{align}
where $s\equiv\frac{\mu_1}{\mu_2}$ and $u\equiv g^2_\mathrm{R}(\mu_1)$.
A common and convenient choice is to take successive scales at fixed ratio $s=2$:
\begin{align}
\sigma_{T}(u)\equiv\sigma_{T}(2,u)=
      \exp  \left \{
            \int\limits_{g_\mathrm{R}(\mu)}^{g_\mathrm{R}(\mu/2)} \, { d\mathnormal{g}} \,{
            \frac{\gamma(\mathnormal{g})}{\beta(\mathnormal{g})}} \right \} \,.
\end{align}
On the lattice, the  scale evolution can be studied non-perturbatively as a finite size scaling,
with the renormalization scale identified as the inverse of the lattice size $\mu = \frac{1}{L}$:
\begin{alignat}{3}
&{\mu = \frac{1}{L}}, \quad u\equiv {g_\mathrm{R}}^2(L)&\qquad&\\
&\sigma(u)=\lim_{a\to 0} \Sigma(u,a/L)   && \Sigma(u,a/L)={g_\mathrm{R}}^2(2L)\\
&\sigma_{T}(u)=\lim_{a\to 0} \Sigma_{T}(u,a/L) && \Sigma_{T}(u,a/L)=\frac{Z_{T}(g_0^2,a/2L)}{Z_{T}(g_0^2,a/L)}\,,
\end{alignat}
where $a$ is the lattice spacing.
The renormalization constants $Z_{T}(g_0^2,a/L)$  are defined imposing renormalization conditions on the correlation functions, as shown in eqs.~(\ref{eq:renconde},\ref{eq:rencondm}) of the next section.   

Our actual RG flow materializes in a sequence of many lattices.  
We used the same gauge configurations generated  by the ALPHA collaboration for the determination of the quark mass running (see  \cite{Campos:2018ahf} for details of the simulations).
They refer to $N_f=3$ massless Wilson-clover fermions with Schr\"odinger Functional (SF) boundary conditions. 
The simulation parameters correspond to a RG evolution from 
 an hadronic scale $\mu_{had}$ of about $200~\rm{MeV}$ to a perturbative scale $\mu_{pt}$ around $128~\rm{GeV}$. 
The peculiarity  of this flow is the change of schemes at the intermediate scale 
$\mu_0/2 \sim 2~\rm{GeV}$: in the high energy region the running coupling is defined in the SF sheme $(g_\mathrm{R}={g}_{SF})$~\cite{Luscher:1992an,Sint:1993un,DallaBrida:2016uha},
while in the low energy region it is defined in the gradient flow (GF) scheme $(g_\mathrm{R}={g}_{GF})$~\cite{DallaBrida:2016kgh,Fritzsch:2013je}:
\begin{center}
 \begin{tikzpicture}
 \coordinate (C);
\draw[->, line width=0.5mm, black]  ($(C)+(0.0,-2.0)$) to node {}($(C)+(8.0,-2.0)$);
\node at ($(C)+(5.5,-1.8)$) [above] {SF scheme};
\node at ($(C)+(2.5,-1.8)$) [above] {GF scheme};
\node at ($(C)+(8.5,-2.0)$) [left] {$\mu$};
\draw[line width=0.5mm, black]  ($(C)+(1.0,-1.9)$) to node[above] {$\mu_{had}$}($(C)+(1.0,-2.1)$);
\draw[line width=0.5mm, black]  ($(C)+(1.0,-1.9)$) to node[below] {$\approx \, 200 \mathrm{MeV}$}($(C)+(1.0,-2.1)$);
\draw[line width=0.5mm, black]  ($(C)+(7.0,-1.9)$) to node[above] {$\mu_{pt}$} ($(C)+(7.0,-2.1)$);
\draw[line width=0.5mm, black]  ($(C)+(7.0,-1.9)$) to node[below] {$\approx \, 128 \mathrm{GeV}$} ($(C)+(7.0,-2.1)$);
\draw[line width=0.5mm, black]  ($(C)+(4.0,-1.9)$) to node[above] {$\mu_0/2$} ($(C)+(4.0,-2.1)$);
\draw[line width=0.5mm, black]  ($(C)+(4.0,-1.9)$) to node[below] {$\approx \, 2 \mathrm{GeV}$} ($(C)+(4.0,-2.1)$);
\end{tikzpicture}
\end{center}
We impose the same definition of $Z_{T}(g_0^2,a/L)$  at all scales, which implies that the anomalous dimension, which is a different function of 
${g}^2_{SF}$ and of ${g}^2_{GF}$, has the same value at a given renormalisation scale $\mu$:
\begin{align}
\gamma(\mu)=\gamma_{SF}(g^2_{SF}(\mu))=\gamma_{GF}(g^2_{GF}(\mu))\,.
\end{align}
%%%
\section{$\chi$SF Chirally Rotated Schr\"odinger Functional}
In our study we adopt a mixed action approach (see also \cite{Campos:2019nus,Plasencia:2021gjp}): 
while the sea quarks obey the standard SF boundary conditions, for the valence quarks we impose $\chi$SF boundary conditions.
In the continuum and chiral limit, the  SF and $\chi$SF  setups are equivalent, being connected by a chiral flavor transformation~\cite{Sint:2010eh}:
\begin{align}
\small R=\left.\exp\left(i\dfrac{\alpha}{2}\gamma_5\tau^3\right)\right\vert_{\alpha=\pi/2}\qquad
\begin{cases}
\psi & \to \psi^\prime= R\psi\\
\bar \psi & \to \bar \psi^\prime = \bar \psi R
\end{cases}\\
P_\pm \equiv \tfrac{1}{2}(1\pm \gamma_0) \to {Q_\pm \equiv \tfrac{1}{2}(1\pm i\gamma_0\gamma_5\tau^3)} \,,
\end{align}
where $P_\pm$ and $Q_\pm$ are the SF and the $\chi$SF projectors acting on fermionic fields at the boundaries.

At finite lattice spacing however, $\chi$SF breaks the parity-flavor symmetry ${\cal P}_5=i\gamma_0\gamma_5\tau^3$, which is recovered by introducing an extra boundary countertem of dimension 3 with coefficient $\zf$.
The parameters $\zf$ and the bare mass $m_0$ must be  tuned non-perturbatively 
to their critical values in order to restore parity-flavor and chiral symmetries up to discretisation effects.  In practice the two tunings can be done independently, so we  inherited the value of the critical hopping parameter $\kappa$ from the SF simulations~\cite{Campos:2018ahf}, while we fixed $\zf$ imposing the vanishing of  $\ga^{ud}$,  a ${\cal P}_5$-odd correlation function:
\begin{align}
\begin{cases}
m = \frac{\tilde\partial_0 f_{{\rm A},I}^{ud}(x_0)}{
2f_{\rm P}^{ud}(x_0)}\big |_{x_0=L/2} = 0 
&m_{cr} \text{ tuning } \\
\ga^{ud}(x_0)\big |_{x_0=L/2}     = 0 & 
 \zf \text{ tuning.}
\end{cases}
\end{align}
Here $m$ stands for the SF-PCAC quark mass, $f_{{\rm A},I}$ and $f_{{\rm P}}$ are the usual SF correlation functions of the improved axial current and the  pseudoscalar density,
 while  $\ga^{ud}$ is the $\chi$SF correlation function involving the axial current with flavors $u,d$. 
See Table \ref{tab:sfchisf} for a brief overview of the correlation functions. Some details of our tuning procedure can be found in ref.~\cite{Plasencia:2021gjp}

A boundary improvement counterterm proportional to the coefficients $d_s$ is also needed  in order to cancel $O(a)$ discretisation effects originating at the time borders.

Once these requirements are fulfilled, the argument of automatic $O(a)$ improvement is achieved in $\chi$SF~\cite{Sint:2010eh,Sint:2010xy}: the ${\cal P}_5$-even  correlation functions 
receive  corrections only at second order in the lattice spacing, whereas the ${\cal P}_5$-odd ones are pure lattice artefacts:
\begin{align}
&g_{even} = g_{even}^{\rm continuum} + O(a^2) \\
&g_{odd} = O(a)\,.
\end{align}
This  property  turns out to be particularly advantageous for the tensor operator, because $O(a)$ improvement does not require mixing with bulk counterterms  in the correlation functions. For $\lt$ with flavor combination ${ud}$, for example, 
the improvement coefficient $c_T$ is irrelevant, since the vector correlation function $\lv$, being   ${\cal P}_5$-odd, is $O(a)$. Therefore the Symanzik correction, being $O(a^2)$, may be dropped:
  \begin{align}
&\qquad T_{\mu \nu}^{\rm I}=T_{\mu \nu} + { c_{\rm T}(g_0^2)\; a\,(\tilde{\partial}_{\mu}V_{\nu} - \tilde{\partial}_{\nu}V_{\mu})}\,,\\
&\qquad \lt^{ud, {\rm I}}=\lt^{ud} + \cancel{ c_{\rm T}(g_0^2)\; a\,\tilde{\partial}_{0}\lv^{ud} }\,.
\end{align}
The rich variety of correlation functions is an interesting feature of $\chi$SF, offering the possibility of several definitions of $Z_{\rm T}$, for example through the renormalization condition 
on the electric tensors as well as on the magnetic ones:
  \begin{align}
& \qquad {Z_{\rm T}}(g_0,a/L)\frac{ {\lt^{ud} (L/2)}_{\rm re}}{\sqrt{l_1^{ud}}} =
  \left.\frac{ {\lt^{ud} (L/2)}_{\rm re}}{\sqrt{l_1^{ud}}}\right|_{\text{Tree Level}}\qquad \text{``electric'' tensor $T_{0 k}$} 
\label{eq:renconde}\\
& \qquad {Z_{\rm T}}(g_0,a/L)\frac{ {\ltt^{uu^\prime} (L/2)}_{\rm im}}{\sqrt{l_1^{uu^\prime}}} =
  \left.\frac{ {\ltt^{uu^\prime} (L/2)}_{\rm im}}{\sqrt{l_1^{uu^\prime}}}\right|_{\text{Tree Level}}\qquad \text{``magnetic'' tensor $\tilde T_{0k}= -\frac12\varepsilon_{0kij} T_{ij}$}\,.
\label{eq:rencondm}
\end{align}
The subscripts $_{\rm re},\; _{\rm im}$ here denote the real and the imaginary parts.
Both relations correspond to the same SF renormalization condition in the continuum. In fact, 
the continuum equivalence of the SF and $\chi$SF setups implies universality relations among correlation functions:
  \begin{alignat}{4}
    \kt  &=&\,i\ltt^{uu'} &=&\,-i\ltt^{dd'} &=&\,   \lt^{ud} &= \,\phantom{i} \lt^{du},
   \label{eq:dictkt}\\
     k_1  &=&\,  l_1^{uu'} &=&\, l_1^{dd'} &=&\,  l_1^{ud} &=\, l_1^{du}.
  \label{eq:dictk1}
 \end{alignat}
\begin{table}
\begin{align*}
&
\qquad \text{\scriptsize flavors} \qquad \qquad  \scriptstyle f_1f_2 = u,d,u^\prime, d^\prime\\
&
  \qquad \text{\scriptsize bulk operators}  \qquad  \scriptstyle X=V_0,A_{0},S,P \quad 
 \scriptstyle Y_{k}=V_{k},A_k,T_{k0},\widetilde{T}_{k0}\\
 & \begin{array}{lll} 
 \toprule
 &\text{SF}&\color{red}\text{$\chi$SF}\\
 &\scriptstyle f_{\rm X}(x_{0})=-{1\over2}\big\langle X^{f_{1}f_{2}}(x)\mathcal{O}_{5}^{f_{2}f_{1}}\big\rangle
 &\scriptstyle g_{\rm X}(x_0) = -\frac12\big\langle X^{f_1f_2}(x){\color{red}{\cal Q}_{5}^{f_2f_1}}\big\rangle\\
  &\scriptstyle k_{\rm Y}(x_{0})=-{1\over6}\sum_{k=1}^{3}\big\langle Y_{k}^{f_{1}f_{2}}(x)\mathcal{O}_{k}^{f_{2}f_{1}}\big\rangle
 &\scriptstyle l_{\rm Y}(x_0) = -\frac16\sum_{k=1}^3\big\langle Y_k^{f_1f_2}(x){\color{red}{\cal Q}_{k}^{f_2f_1}}\big\rangle \\ \\
  &\scriptstyle f_{1}= - \frac12 \big\langle {\cal O}_{5}^{f_1f_2} {\cal O}_{5}^{'f_2f_1}\big\rangle
&\scriptstyle g_{1} =
 - \frac12 \big\langle {\color{red}{\cal Q}_{5}^{f_1f_2} {\cal Q}_{5}^{'f_2f_1}}\big\rangle\\
  &\scriptstyle k_{1}= - \frac16\sum_{k=1}^3\big\langle {\cal O}_{k}^{f_1f_2} {\cal O}_{k}^{'f_2f_1}\big\rangle
 &\scriptstyle l_{1}= - \frac16\sum_{k=1}^3
           \big\langle {\color{red}{\cal Q}_{k}^{f_1f_2} {\cal Q}_{k}^{'f_2f_1}}\big\rangle\\ \\
& \scriptstyle \mathcal{O}_{5}^{f_{1}f_{2}} =a^{6}\sum_{\mathbf{y,z}}\overline{\zeta}_{f_{1}}(\mathbf{y})P_{+}\gamma_{5}\zeta_{f_{2}}(\mathbf{z})
&  \scriptstyle {\color{red}\mathcal{Q}_{5}^{uu'}} =a^{6}\sum_{\mathbf{y,z}}\overline{\zeta}_{u}(\mathbf{y})\gamma_{0}\gamma_{5}{\color{red} Q_{-}}\zeta_{u'}(\mathbf{z})\\
 & \scriptstyle \mathcal{O}_{k}^{f_{1}f_{2}} =a^{6}\sum_{\mathbf{y,z}}\overline{\zeta}_{f_{1}}(\mathbf{y})P_{+}\gamma_{k}\zeta_{f_{2}}(\mathbf{z})
 & \scriptstyle {\color{red} \mathcal{Q}_{k}^{uu'}} =a^{6}\sum_{\mathbf{y,z}}\overline{\zeta}_{u}(\mathbf{y})\gamma_{k}{ \color{red} Q_{-}}\zeta_{u'}(\mathbf{z})\\
& \vdots & \color{red} \vdots\\
&P_\pm \equiv \tfrac{1}{2}(1\pm \gamma_0) & \color{red} {Q_\pm \equiv \tfrac{1}{2}(1\pm i\gamma_0\gamma_5)} \\
 \toprule
\end{array}
\end{align*}
\caption{Brief overview of correlation functions in SF and $\chi$SF setups.}
\label{tab:sfchisf}
\end{table}

%%%
\section{Results}
\begin{figure}[!ht]
\centering
{\includegraphics[width=0.85\textwidth]{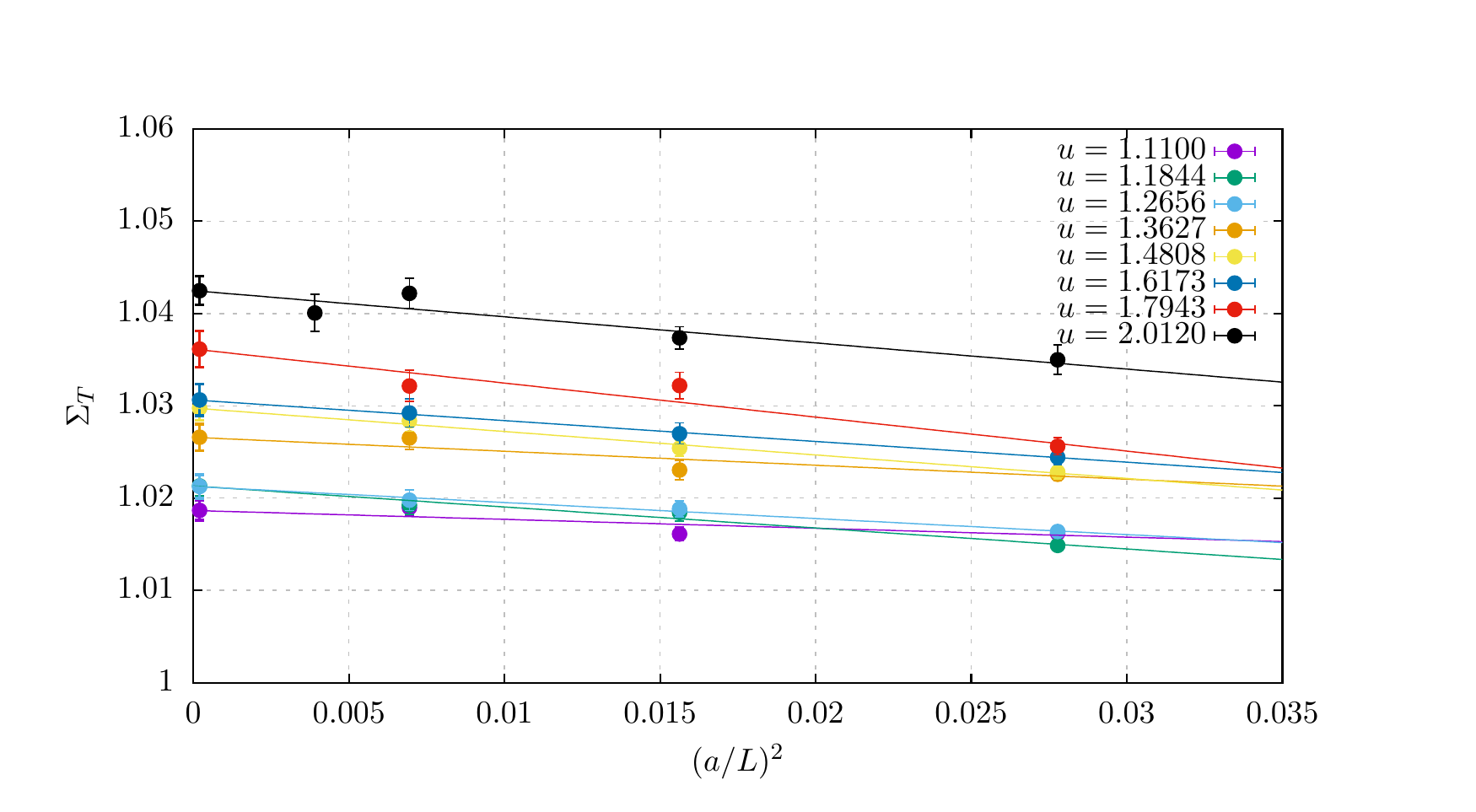}}
\caption{The step scaling function $\Sigma_{\rm T}(u,a/L)$ in the high energy region}
\label{fig:SigmaTa2}
\end{figure}

\begin{figure}[!ht]
\centering
\begin{tabular}{cc}
{\includegraphics[width=0.5\textwidth]{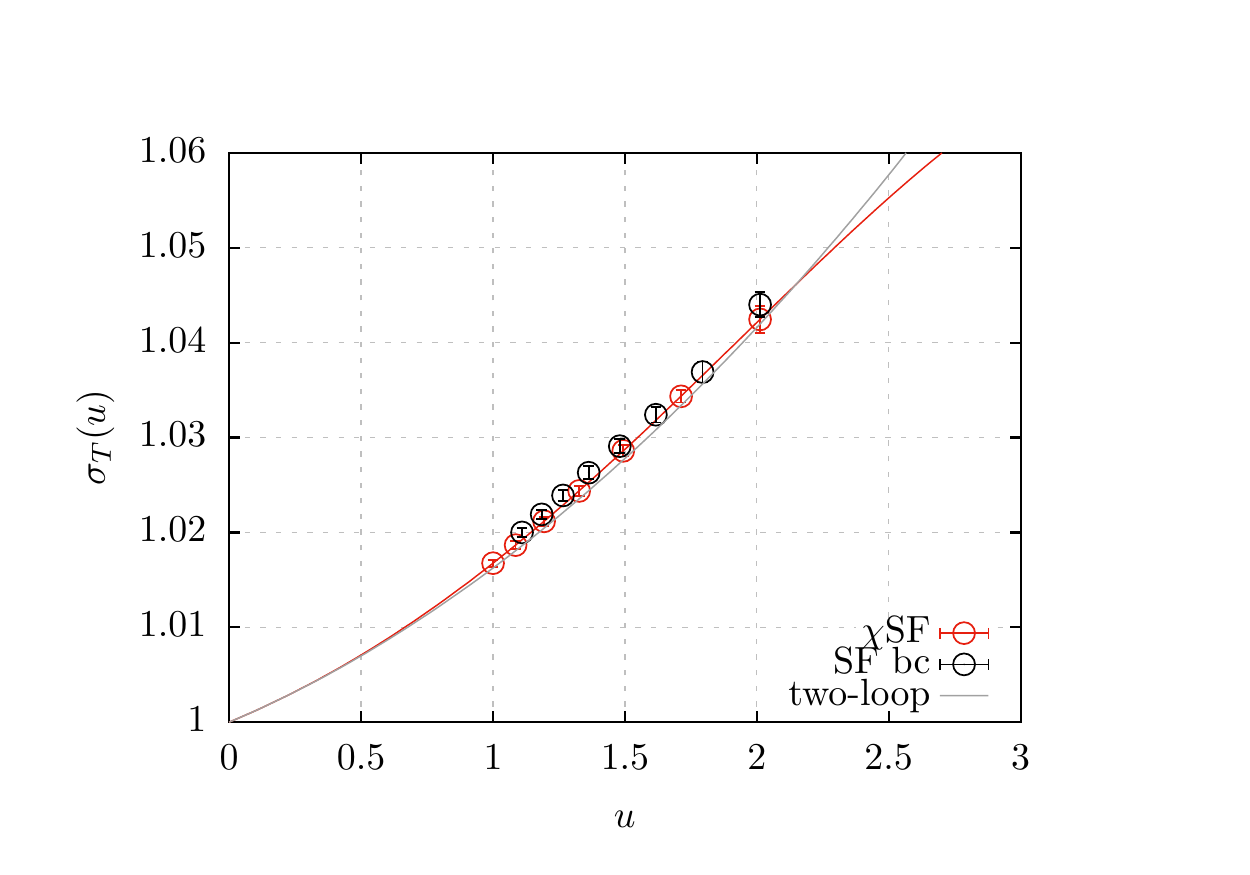}}
{\includegraphics[width=0.5\textwidth]{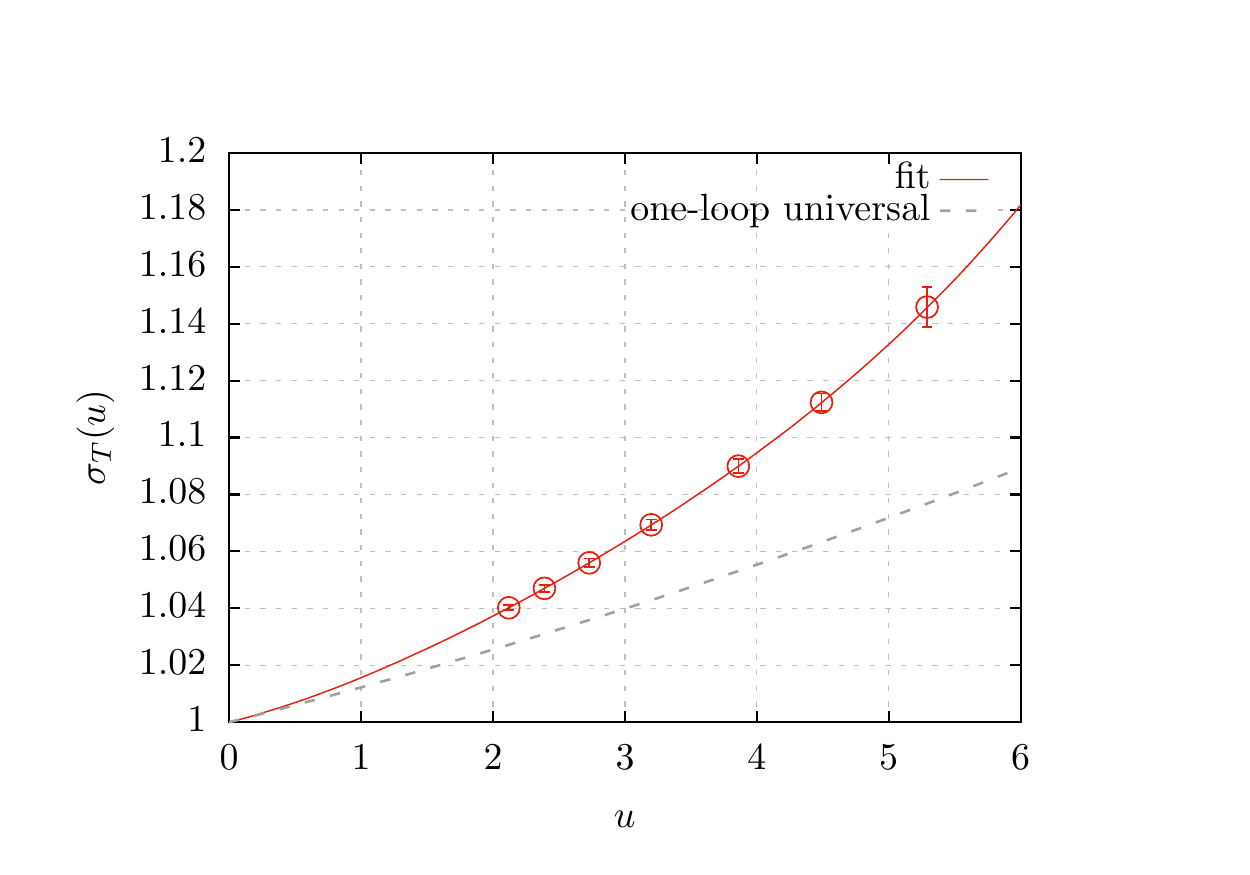}}
\end{tabular}    
\caption{The continuum step scaling function $\sigma_\mathrm{T}(u)$ at  high energies (on the left) and at low energies (on the right).
The two energy regions correspond to different definitions of the running coupling:  $u={g}^2_{SF}$  ~\cite{Luscher:1992an,Sint:1993un,DallaBrida:2016uha} and
 $u={g}^2_{GF}$~\cite{DallaBrida:2016kgh,Fritzsch:2013je}.}
\label{fig:sigmaTu}
\end{figure}

\begin{figure}
\centering
{\includegraphics[width=0.6\textwidth]{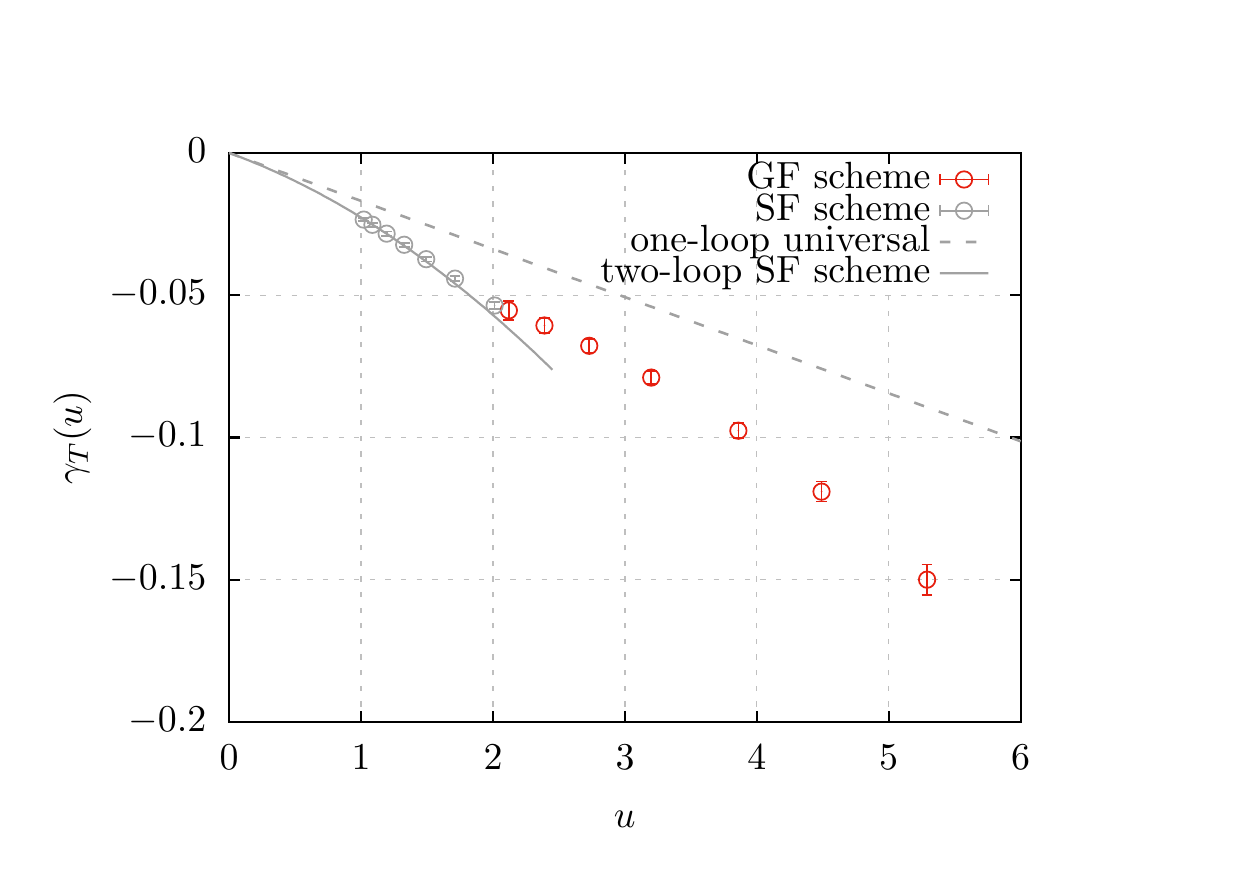}}
\caption{The anomalous dimensions $\gamma_{SF}(u)$ and $\gamma_{GF}(u)$.
The two schemes SF and GF correspond to different definitions of the  running coupling:  $u={g}^2_{SF}$  ~\cite{Luscher:1992an,Sint:1993un,DallaBrida:2016uha} and
 $u={g}^2_{GF}$~\cite{DallaBrida:2016kgh,Fritzsch:2013je}.}
\label{fig:gammaTu}
\end{figure}

Our preliminary results are based on the determination of $Z_{\rm T}$ from eq.~(\ref{eq:renconde}).
We obtain the continuum limit of the tensor lattice step scaling function $\Sigma_{\rm T}(u,a/L) $ performing global fits of the data at different couplings and lattice spacings:
\begin{align}
\Sigma_{\rm T}(u,a/L) &= \frac{Z_{T}(g_0^2,a/2L)}{Z_{T}(g_0^2,a/L)}\\
&= \sigma_{\rm T}(u) + \rho_{\rm T}(u)\left(\frac{a}{L}\right)^2 \,,
\end{align}
with  $\sigma_{\rm T}(u) $ and  $\rho_{\rm T}(u) $ parameterised by polynomials.
Fig.~\ref{fig:SigmaTa2}  shows  $\Sigma_{\rm T}$
as a function of $(a/L)^2$, and parametrized by the coupling $u$ (in different colours) 
for the high energy region.  The relative continuum limit is illustrated in the left panel of Fig.~\ref{fig:sigmaTu}.
Our data (red circles) are in agreement with perturbation theory at two loops (gray line)~\cite{Pena:2017hct}
and with the data obtained in a purely SF setup (circles in black)~\cite{Chimirri:2019xsv,Fabian}. This demonstrates  the continuum SF-$\chi$SF universality.
The results at low energies  are shown in the right panel of the same figure. 
We have extracted  the anomalous dimension $\gamma$, relying on the formula:
\begin{align}
\sigma_\mathrm{T}(u)  = \exp   \bigg\{ \int\limits_{\sqrt u}^{\sqrt {\sigma(u)}} \, d\mathnormal{g} \, 
              \frac{\gamma(\mathnormal{g})}{\beta(\mathnormal{g})} \bigg \} \,.
\end{align}
In Fig.~\ref{fig:gammaTu}  we then finally show our preliminary results for $\gamma_{SF}$ and $ \gamma_{GF}$, over the full range of couplings available.  

%%%
\section{Conclusions}
We have presented  prelimirary results for the RG running of flavor non-singlet tensor operator in $N_f=3$ QCD, using the gauge configurations generated by the ALPHA collaboration~\cite{Campos:2018ahf}.
The data span, in a fully non-perturbative way, a range of energies of about three orders of magnitude, 
going from hadronic to electro-weak scales. We obtained the anomalous dimension of the tensor operator, aiming to
complete the computation of the non-perturbative RG running of all dimension 3 bilinear operators.
\section{Acknowledgements}
We wish to thank Patrick Fritzsch, Carlos Pena, David Preti, and Alberto Ramos for their help. This work  is partially supported by INFN and CINECA, as part of research project of the QCDLAT INFN-initiative. 
We acknowledge the Santander Supercomputacion support group at the University of Cantabria which provided access to the Altamira Supercomputer at the Institute of Physics of Cantabria (IFCA-CSIC).
We also acknowledge support by the Poznan Supercomputing and Networking Center (PSNC) under the project with grant number 466.
AL acknowledges support by the U.S.\ Department of Energy under grant number DE-SC0015655.

\end{document}